In press at *Nature Geoscience*

# The future life span of Earth's oxygenated atmosphere

Kazumi Ozaki[1,2*] and Christopher T. Reinhard[2,3,4]

[1]Department of Environmental Science, Toho University, Funabashi, Chiba 274-8510, Japan

[2]NASA Nexus for Exoplanet System Science (NExSS)

[3]School of Earth and Atmospheric Sciences, Georgia Institute of Technology, Atlanta, GA 30332, USA

[4]NASA Astrobiology Institute, Alternative Earths Team, Riverside, CA, USA

*Correspondence to: kazumi.ozaki@sci.toho-u.ac.jp

**Abstract:**

Earth's modern atmosphere is highly oxygenated and is a remotely detectable signal of its surface biosphere. However, the lifespan of oxygen-based biosignatures in Earth's atmosphere remains uncertain, particularly for the distant future. Here we use a combined biogeochemistry and climate model to examine the likely timescale of oxygen-rich atmospheric conditions on Earth. Using a stochastic approach, we find that the mean future lifespan of Earth's atmosphere, with oxygen levels more than 1% of the present atmospheric level, is $1.08 \pm 0.14$ billion years ($1\sigma$). The model projects that a deoxygenation of the atmosphere, with atmospheric $O_2$ dropping sharply to levels reminiscent of the Archaean Earth, will most probably be triggered before the inception of moist greenhouse conditions in Earth's climate system and before the extensive loss of surface water from the atmosphere. We find that future deoxygenation is an inevitable consequence of increasing solar fluxes, whereas its precise timing is modulated by the exchange flux of reducing power between the mantle and the ocean–atmosphere–crust system. Our results suggest that the planetary carbonate–silicate cycle will tend to lead to terminally $CO_2$-limited biospheres and rapid atmospheric deoxygenation, emphasizing the need for robust atmospheric biosignatures applicable to weakly oxygenated and anoxic exoplanet atmospheres and highlighting the potential importance of atmospheric organic haze during the terminal stages of planetary habitability.

***Introduction***

The field of exoplanet astronomy is on the verge of the first detailed characterization of the atmospheres of extrasolar planets, with heightened expectations for the potential detection of spectroscopic indicators of habitability, and possibly life, using the next generation of space and ground-based telescopes[1-3]. However, a robust framework for remote life detection requires a detailed understanding of the factors that lead to the emergence and long-term maintenance of atmospheric biosignatures, including a full understanding of potential atmospheric composition in the broader context of planetary biogeochemistry[4]. Although numerous potential atmospheric biosignatures have been proposed[5], molecular oxygen ($O_2$) and its photochemical by-product ozone ($O_3$) remain at the forefront of approaches toward remote life detection due to their strong absorption at visible and near-infrared wavelengths, the importance of biology in controlling their atmospheric abundances, and the potential for homogeneity in abundance throughout the atmospheric column[4,6]. As a result, significant effort has been directed toward understanding atmospheric oxygenation on habitable worlds and the factors that might lead to abiotic oxygen-rich atmospheres[7-9].

Earth's photosynthetic biosphere currently supports large gross fluxes of $O_2$ to the ocean-atmosphere system—roughly $9\times10^{15}$ mol $O_2$ $y^{-1}$ (ref.[10])—with the result that the modern atmosphere is 20.5% $O_2$ by volume. However, the presence of oxygenic phototrophs on its own may not be enough to maintain a strongly oxygenated atmosphere. Indeed, it is generally thought that the abundance of $O_2$ in Earth's atmosphere has been well below that of today for most of Earth's history[11], and that the modern atmospheric $O_2$ abundance developed only after the emergence of land plants[12,13], which have evolved to accelerate the geochemical cycles of bioessential elements (most importantly phosphorus) in Earth surface environments[14,15]. Thus, charting the history of planetary oxygenation after the evolution of oxygenic photosynthesis requires an understanding of how biospheric productivity (and the subsequent burial of reduced materials into sediments) changes in response to evolving environmental conditions.

Previous work on the future lifespan of Earth's biosphere has focused principally on the links between secular changes in solar luminosity, the stability of the carbonate-silicate geochemical cycle, and the loss of surface water to space[16-20]. One robust result that emerges from this theoretical framework is a continuous decrease in the abundance of atmospheric $CO_2$ through time, as steadily increasing solar luminosity inexorably drives down the abundance of atmospheric $CO_2$ that is required to balance rates of crustal weathering with volcanic $CO_2$ output[16,17,21]. However, a number of models suggest that in spite of this thermal buffering mechanism Earth will enter the "moist greenhouse" climate state at some point within the next ~2 billion years (Ga)—a regime in which elevated atmospheric temperatures allow Earth's stratosphere to become moist, leading to permanent water loss from Earth's surface[16-19,22,23]. In addition, this drawdown of atmospheric $CO_2$ may result in a fundamental shift to a photosynthetic biosphere that is $CO_2$-limited—for instance, some models predict that terrestrial $C_3$ plants will cease to be viable at Earth's surface less than ~500 Myr into Earth's future[16]. If true, this might place a long-term physical limit on the ability of photosynthetic biospheres to maintain high levels of atmospheric oxygen, giving rise to a fundamental trade-off between long-term stellar evolution, the geologic carbon cycle, and the intrinsic timescale of atmospheric oxygenation.

Here, we examine the future lifespan of Earth's oxygenated atmosphere using an Earth system model of biogeochemistry and climate (**Fig. 1**; see Methods). Assessing the lifespan of Earth's oxygenated atmosphere requires quantitative models that include a complete treatment of oxygen-related biogeochemical processes at the planetary surface over geologic time, as well as mechanistic linkages between surface environments and the planetary interior. Our model builds upon previous similar Earth system models[12] that track the coupled carbon, oxygen, phosphorus, and sulphur cycles in the exogenic (i.e., atmosphere, ocean, and crust) system (**Fig. 1**). However, we extend this framework by incorporating a global biogeochemical methane ($CH_4$) cycle, including a range of key biological metabolisms, parameterized atmospheric photochemistry within the $O_2$-$O_3$-$CH_4$ system[24,25], and the radiative impact of $CH_4$ on global energy balance[26].

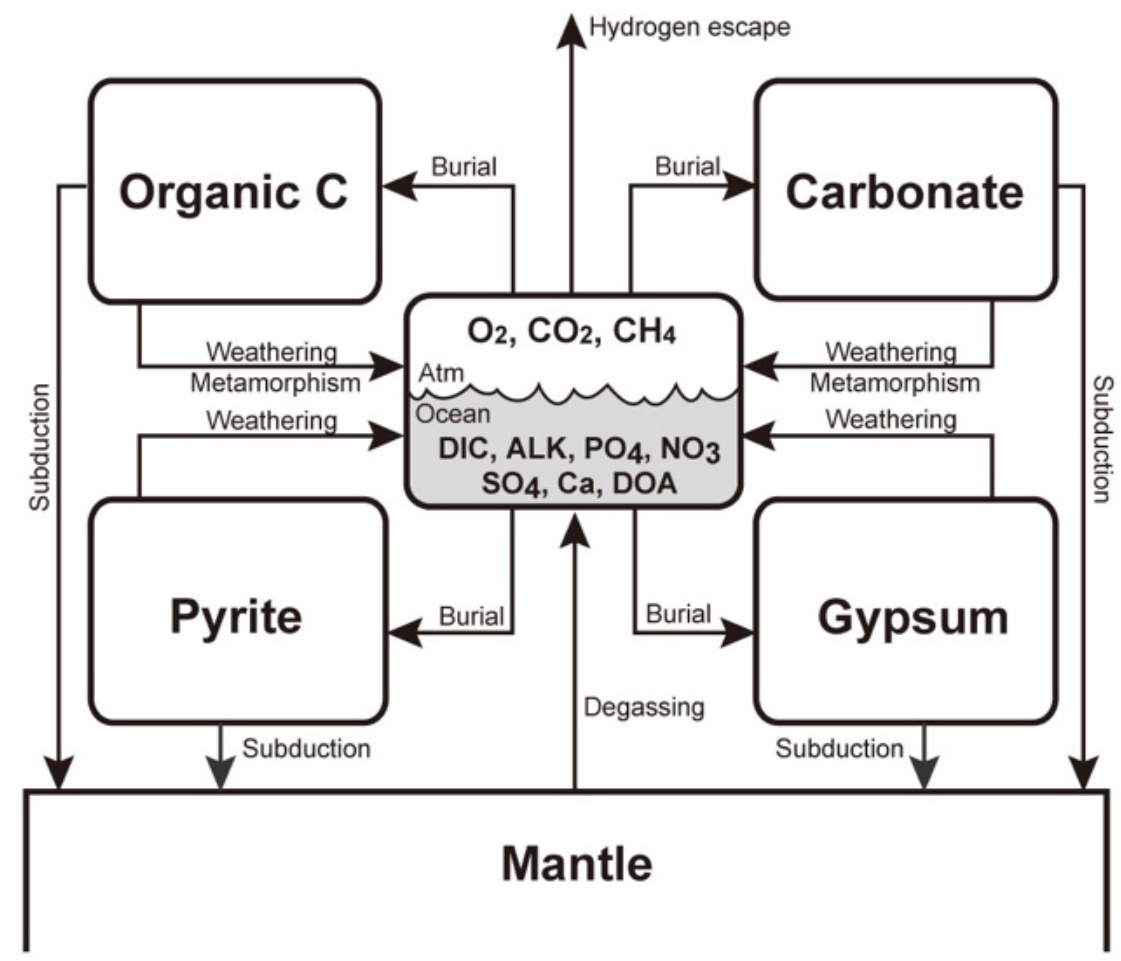

**Fig. 1. Schematic model structure.** Boxes denote reservoirs, whereas arrows denote flux terms. The model tracks the major reservoirs and transfer fluxes within the surface carbon (C), sulphur (S), oxygen (O), and phosphorus (P) cycles, along with a comprehensive treatment of ocean biogeochemistry, and long-term transfers between the crust-ocean-atmosphere system and the mantle. DOA = degree of anoxia.

Our model is also the first of its type to explicitly evaluate redox exchange between the crust and mantle, allowing a comprehensive evaluation of the planetary processes controlling atmospheric $O_2$ levels over geologic time. The model is designed to capture the major components of Earth system biogeochemistry and climate, but is abstracted enough to allow for large model ensembles that can be run for billions of years of model time. We employ a stochastic approach in which we draw randomly from a wide range of potential values for parameters controlling the major geophysical boundary conditions of the model, including variability (cycle and amplitude) in outgassing rates and erosional forcing (see Methods and Extended Data Figs. 1-3), and initialize the model at 600 Ma. We then repeatedly run the model forward in time until the present day (n =~400,000), and subsample model runs (n = 4,787) that roughly reproduce modern atmospheric composition ($pO_2$ = 21±1% and $pCO_2$ = 300±150 ppmv), global average surface temperature (288±2K), and seawater sulphate ($SO_4^{2-}$) levels ($SO_4^{2-}$ = 25±15 mM). The simulations from this subsampled ensemble are then run forward in time in order to estimate the future lifespan of Earth's oxygenated atmosphere.

***Solar brightening promotes atmospheric deoxygenation***

Our default analysis (**Fig. 2** and Extended Data Figs. 4-8) indicates that atmospheric $O_2$ levels will decrease substantially in Earth's long-term future (**Fig. 2a**). Whilst the magnitude and timing of this trajectory show a certain degree of uncertainty, the model-projected decrease in atmospheric $O_2$ abundance is robust and suggests that the future lifespan of Earth's oxygenated atmosphere will be less than 1.5 Gyr. Specifically, when we define the future lifespan as the period within which atmospheric $O_2$ remains above a given threshold value, the lifespan of atmospheric $O_2$ levels greater than 1% of the present atmospheric level (PAL)—$\tau_{1\%PAL}$—is estimated at 1.08±0.14 Gyr (1σ). Increasing the threshold value to 10% PAL results in a broadly similar lifespan of $\tau_{10\%PAL}$ = 1.05±0.16 Gyr (1σ), a consequence of the extremely short timescale over which atmospheric deoxygenation occurs. Several factors are responsible for the secular decrease in atmospheric $O_2$ (see Supplementary Information), most notably increasing surface temperatures and atmospheric $CO_2$ starvation of the terrestrial photosynthetic biosphere driven by solar brightening (**Fig. 2c** and Extended Data Figs. 4-7). Indeed, when we repeat our analysis with a constant solar luminosity we find that no secular $O_2$ trend remains (**Fig. 2a** and Extended Data Fig. 4). As the geologic drivers of the decreasing $CO_2$ trend (solar brightening and mantle cooling) are largely undisputed, the first-order features of our projected atmospheric $O_2$ trajectory should be robust.

Our model further predicts that $CO_2$ limitation of Earth's photosynthetic biosphere should ultimately lead to photochemical destabilization of Earth's oxygenated atmosphere and an abrupt shift in atmospheric $O_2$ abundance to very low values (**Fig. 2a**). Indeed, our model projects an atmospheric composition for Earth's long-term future that is similar in many respects to that of the Archaean Earth prior to the so-called 'Great Oxidation Event'[27]. In particular, atmospheric $O_2$ drops to values that are many orders of magnitude below that of the modern atmosphere ($<10^{-6}$ PAL), while the atmospheric abundance of $CH_4$ increases sharply (Fig. 2b). However, a critical difference from the Archaean Earth system is the prediction of much lower atmospheric $CO_2$ levels in Earth's long-term future (**Fig. 2c**) and thus the potential for greatly elevated $CH_4/CO_2$ ratios (**Fig. 2d**). Such elevated $CH_4/CO_2$ ratios are generally predicted to result in the formation of 'organic haze' in the atmosphere[28-30], with potentially important impacts on climate, atmospheric redox chemistry, and remote life detection. Our results thus suggest that organic haze will be a critical component of Earth's long-term climate stability into the far future, as well as a potentially promising atmospheric biosignature on Earth-like worlds round late main-sequence stars such as the exoplanet candidate Kepler-452b (orbiting a G2 star with an age of ~6±2 Gyr and receiving ~10% more insolation than Earth receives from the Sun today)[31,32].

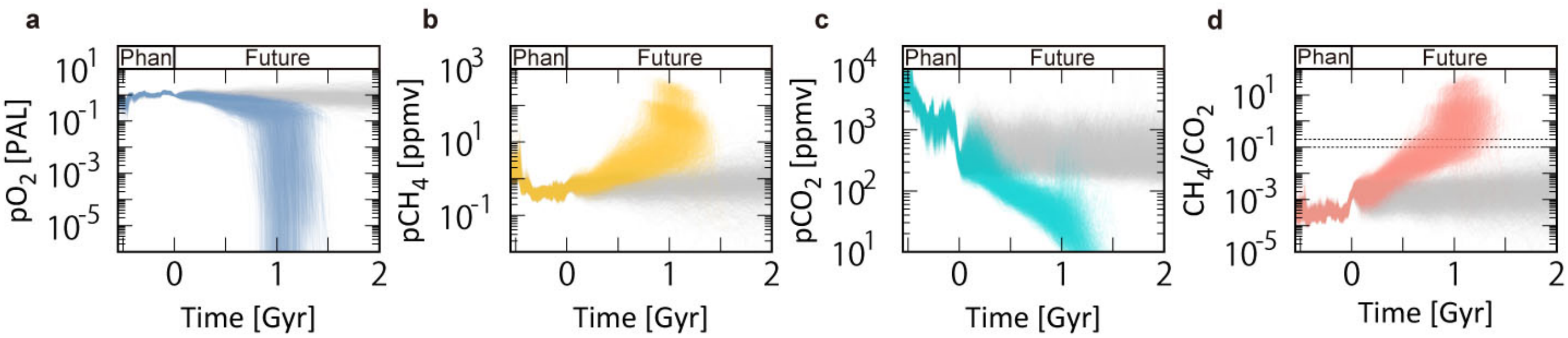

**Fig. 2. Evolution of atmospheric chemistry in our stochastic analysis.** Blue, orange, and green lines represent atmospheric $O_2$ (**a**), $CH_4$ (**b**), and $CO_2$ (**c**) concentrations obtained from our default model ensemble (*n* = 4,787), along with the predicted atmospheric $CH_4/CO_2$ ratios (red lines in **d**). Horizontal dotted lines in **d** represent a range of values between 0.1-0.2, broadly considered to define a threshold for the inception of organic haze formation[28-30]. All panels also show results obtained by running the same model ensemble with constant solar luminosity (grey lines). Phan = Phanerozoic.

***Controls on the timing of atmospheric deoxygenation***

Our default analysis provides the first systematic estimate of the potential lifetime of Earth's oxygenated atmosphere into the far future. However, it is important to note that our default model ensemble includes a robust terrestrial biosphere, which may not be a universal feature of habitable Earth-like planets. To examine the impact of the terrestrial biosphere on the longevity of the $O_2/O_3$-based biosignatures, we performed an additional ensemble analysis in which we remove the terrestrial biosphere from our default successful runs (Extended Data Figs. 4 and 10). As expected, the absence of land plants results in somewhat lower atmospheric $O_2$ levels throughout the course of planetary evolution relative to our default analysis. However, we find that the future lifespan of Earth's remotely-detectable oxygenated atmosphere is similar to that of our default analysis (e.g., $\tau_{1\%PAL}$ = $1.04^{+0.17}_{-0.16}$ Gyr; cf. 1.08±0.14 Gyr for the default simulations) (**Fig. 3a**). This is an important result, as it strongly suggests that the presence or absence of terrestrial biosphere exerts a secondary control on the timescale of the planetary deoxygenation and long-term atmospheric redox state. Although it is difficult to explicitly model uncertainties associated with biological evolution and adaptation in photosynthetic lineages, our statistical approach explores a wide range of parameter space for key biological response functions (see Methods and Extended Data Figs. 1-3), with results suggesting that uncertainty in the parameters governing the response of the biosphere plays a secondary role in controlling the future lifespan of oxygenated atmospheric conditions on Earth (Extended Data Fig. 9).

Our analysis also delineates an important relationship between the future lifespan of oxygenated atmospheric conditions and the long-term average flux of reducing power from the mantle to the surface system, $\Phi_{1\%}$(Red) (**Fig. 3b**). In particular, our model suggests that larger $\Phi_{1\%}$(Red) values tend to yield shorter oxygenated lifespans. Improvements in quantitative estimates of subduction and mantle degassing over time would reduce uncertainty in estimates of the future lifespan of Earth's oxygenated atmosphere, and will be critical for extending our model framework to very different tectonic regimes. In addition, our analysis suggests that oceanic redox chemistry—which ultimately controls oceanic P levels and the redox state and subduction fluxes of major C, S, and Fe species in the exogenic system—would be critical for the long-term trajectory of atmospheric redox state on habitable Earth-like planets (**Fig. 3b** and Extended Data Fig. 4). The lifespans of oxygen-rich atmospheric states on Earth-like planets are thus likely to be modulated by mantle redox state, rates of surface-interior exchange, and their evolution through time.

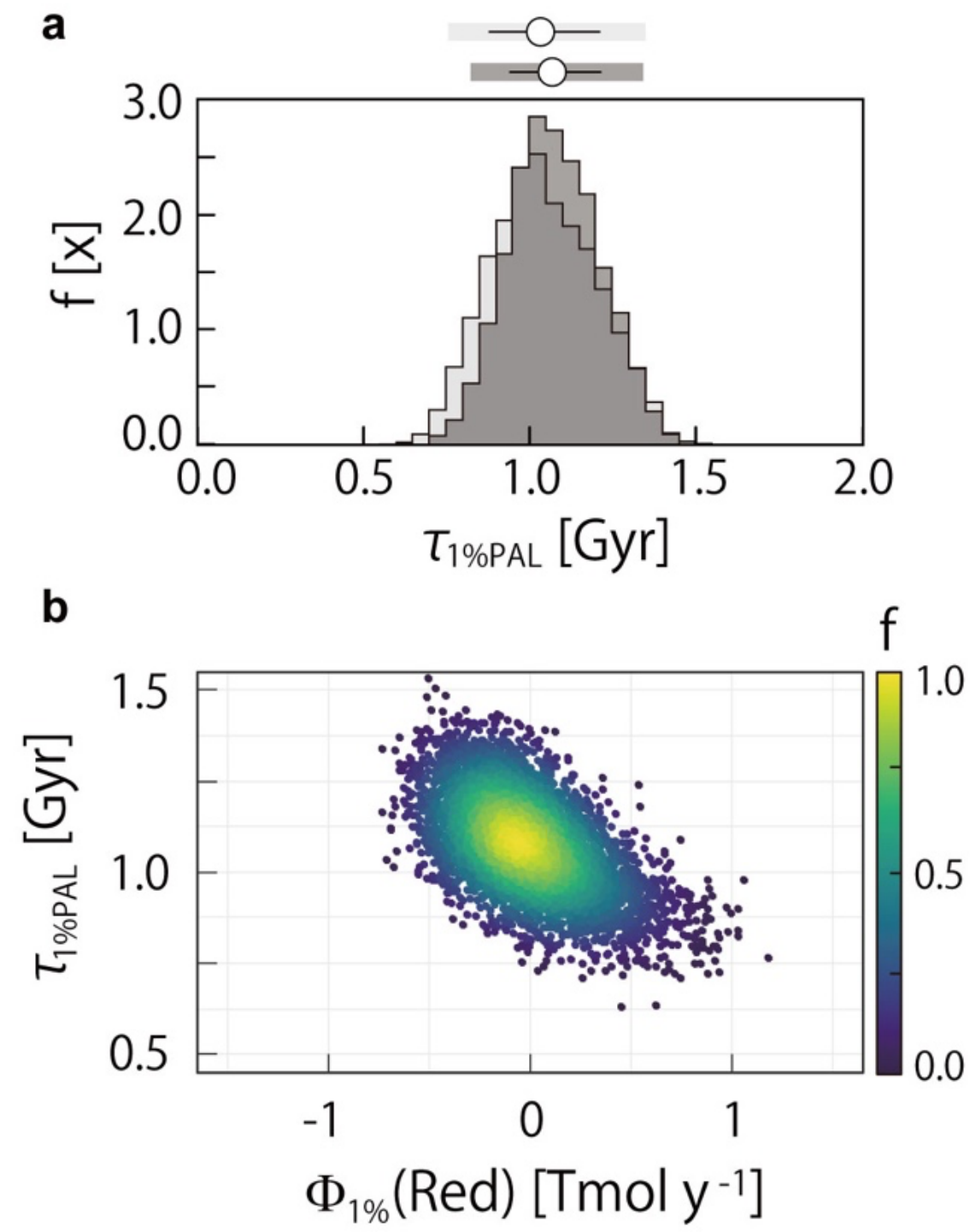

**Fig. 3. The future lifespan of Earth's oxygenated atmosphere.** (**a**) Posterior probability density distributions from our default analysis for Earth systems with (dark grey) and without (light grey) a terrestrial biosphere. Open circles denote median values, while the error bar and shaded region denote 1σ and 95% credible intervals, respectively. (**b**) The future lifespan of Earth's oxygenated atmosphere ($\tau_{1\%PAL}$) as a function of the average net input flux of reducing power from the mantle to the surface system. Positive values indicate net input of reducing power from the mantle to the surface system (see Methods). Each point is a single model run, while the colour scale denotes data density within the overall model ensemble. We note that there are a small number of scenarios in which atmospheric $O_2$ levels do not reach 1%PAL during the simulation, but such scenarios are very rare (~4%) and would rely on a fortuitous combination of parameter values.

It is important to note that there are multiple biogeochemical and climate processes that are not considered in our model that may play a role in constraining the future lifespan of Earth's biosphere and the timing/mode of transition to more reducing atmospheric conditions. In particular, "reverse weathering" (the formation of authigenic silicates in marine sediments, resulting in net $CO_2$ release to the ocean-atmosphere system)[33] could potentially extend the lifespan of oxygenated atmospheric conditions under certain scenarios by prolonging the timescale over which atmospheric $CO_2$ is above the levels expected to result in $CO_2$ limitation of the photosynthetic biosphere[34]. On the other hand, this process would commensurately elevate global temperature in the face of increasing solar luminosity, with a corresponding potential for limitation of biospheric $O_2$ fluxes (Extended Data Fig. 1). The net effect on the future lifespan of Earth's oxygenated atmosphere is uncertain, but is likely of secondary importance to the overall trend presented here (see also Extended Data Fig. 5 for the sensitivity to terrestrial weatherability). In addition, we hypothesize that haze-induced climate cooling[28] could potentially act as a brake on the overall magnitude of atmospheric deoxygenation, or result in the inception of oxygenation/deoxygenation cycles during Earth's terminal habitability. Both of these topics represent important areas of future research.

### *Implications for the search for life on exoplanets*

Our results also have important implications for the search for life on Earth-like planets beyond our solar system (e.g., habitable planets with abundant liquid water at the surface, exposed silicate crust, and a biosphere with oxygenic photosynthesis). From the perspective of planetary evolution, our results imply that atmospheric oxygenation is not a permanent condition on habitable worlds hosting oxygenic photosynthesis, and that only a fraction of Earth's history will be characterized by robustly detectable levels of atmospheric $O_2$ (**Fig. 4**). For example, it is estimated that a G-dwarf stellar flux ~20% greater than that of the present Sun ($S/S_0$ ~ 1.2) is required to achieve a water loss timescale of ~1 Gyr in the moist greenhouse regime[35]. Assuming it will take ~2 Gyr to achieve $S/S_0$ ~ 1.2 (ref.[36]), together with the assumption of surface liquid water by 4.4 Ga[37], yields an estimated habitable lifetime on Earth of ~7.4 Gyr. Our results indicate that direct detection of $O_2$ at visible wavelengths would be challenging for all but ~1.5-2.0 Gyr of this history — or roughly 20-30% of Earth's lifetime as a habitable world[38]. On the other hand, observations of $O_3$ at UV wavelengths[39] could potentially significantly extend this observability timescale (Fig. 4). In any case, our model suggests that a fundamental feature of the carbonate-silicate cycle, considered central to the long-term maintenance of habitable climate conditions on terrestrial planets, may ultimately act to drive atmospheric composition toward an anoxic, hazy state similar to that of the Archaean Earth but with much lower atmospheric $CO_2$. These predictions emphasize the need to better understand the factors regulating redox balance between the planetary interior and surface environments[40-44], and provide additional motivation for implementing a wide wavelength range in future exoplanet characterization missions as a contingency against "false negatives" for remote life detection[3,38].

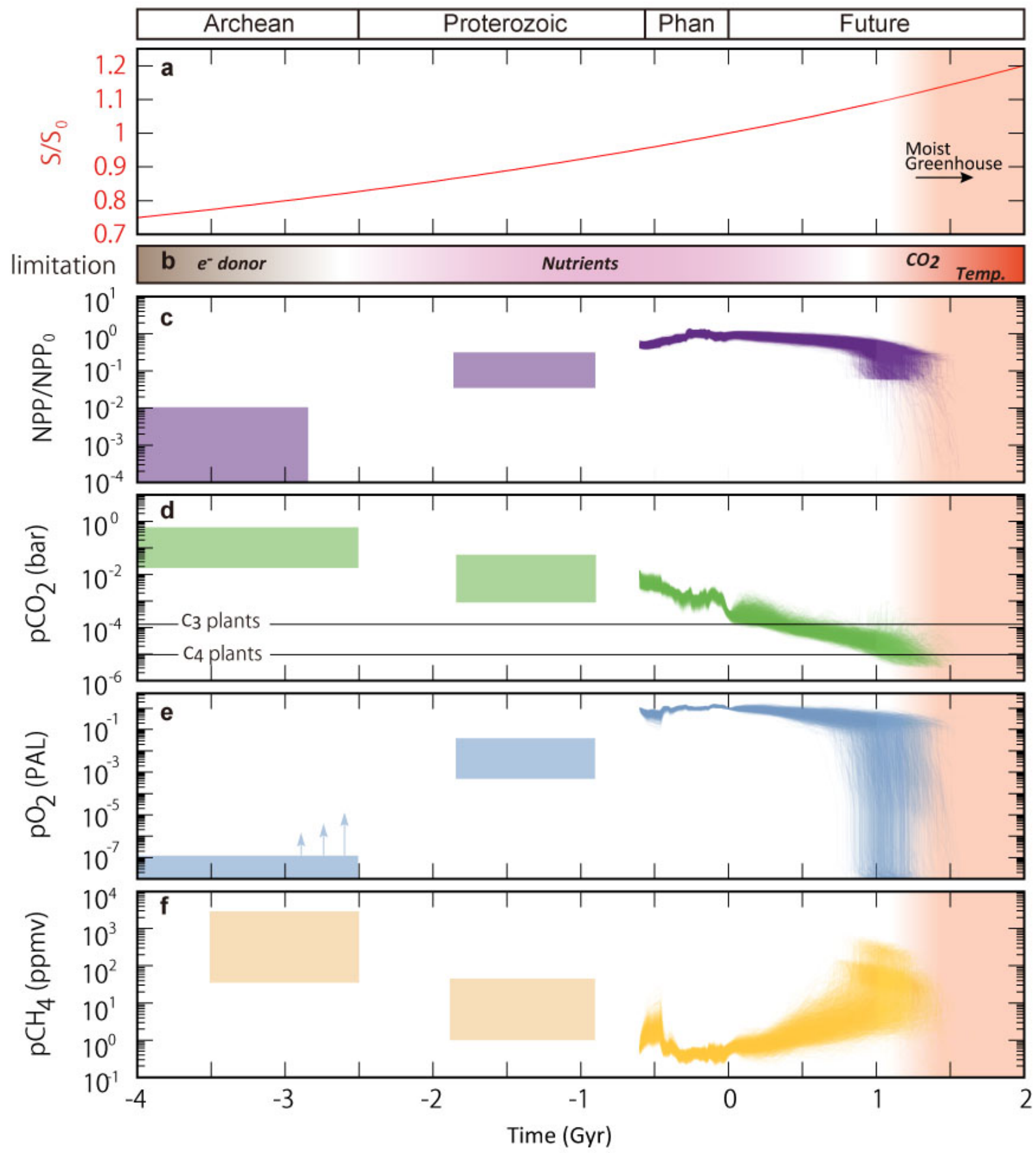

**Fig. 4. The coupled evolution of Earth's biosphere and atmospheric chemistry.** (a) Solar constant (normalized by modern value)[16]. (b) Limiting factor for biospheric activity. (c) Global net primary production (normalized to the modern value)[45,46]. (d) Atmospheric $CO_2$ level (in bar) [47]. (e) Atmospheric $O_2$ level (relative to the present atmospheric level, PAL)[11]. (f) Atmospheric $CH_4$ level (in parts per million by volume, ppmv). The atmospheric $CH_4$ level and global NPP during the mid-Archean and mid-Proterozoic are based on ref.[45,46]. The historical trajectory of biospheric productivity suggests that despite evolving limitations on overall biospheric fertility the biosphere has significantly altered the chemical composition of the atmosphere[45,46,48-50]. Our model, however, demonstrates that Earth's modern biosphere reaches the turning point at which the activity level of the biosphere begins to decline (see also Extended Data Figs. 4-7), and that the availability of inorganic carbon in combination with increased surface temperature will become the major limiting factor for global ecosystems. The orange-red shaded region denotes the inception of the moist greenhouse climate regime.

In sum, our stochastic analysis suggests that the eventual deoxygenation of Earth's atmosphere is a robust outcome of increasing solar luminosity, irrespective of large uncertainties in geophysical/biological boundary conditions, and that the collapse of Earth's oxygenated atmosphere is likely to occur before the inception of moist greenhouse conditions in Earth's climate system although its precise timing is modulated by net input flux of reducing power from the mantle to Earth's surface system. In addition, our model projections of atmospheric chemistry implicate organic haze as a critical factor in regulating Earth's long-term future climate. Evaluating these factors using more sophisticated biogeochemistry and climate models is an important task for future research. Nevertheless, our analysis clearly identifies the first-order response of atmospheric $O_2$ to declining $CO_2$ and biospheric productivity, opening a new perspective on the future evolution of Earth's climate and atmospheric redox chemistry.

**Corresponding author:**

Correspondence to: Kazumi Ozaki, E-mail: kazumi.ozaki@sci.toho-u.ac.jp

**Acknowledgements:** The authors thank E. Tajika, Y. Sekine, S. Kadoya, Y. Watanabe, and E. Schwieterman for helpful discussions. K.O. acknowledges support from the NASA Postdoctoral Program at the NASA Astrobiology Program, administered by Universities Space Research Association under contact with NASA. This work was supported by JSPS KAKENHI Grant Number JP20K04066. C.T.R. acknowledges support from the NASA Astrobiology Institute (Grant 13-13NAI7_2-0027) and the NASA Nexus for Exoplanet System Science (NExSS) (Grant 80NSSC19KO461).

**Author Contributions:** K.O. and C.T.R. designed the study. K.O. constructed the model and performed experiments. K.O. and C.T.R. analysed the results and wrote the paper. All authors discussed and contributed intellectually to the interpretation of the results.

**Competing interests:** The authors declare no competing interests.

## Methods

**Model description.** Our model is based on the Carbon Oxygen Phosphorus Sulphur Evolution (COPSE) model[12,48], the default version of which is described in full elsewhere[12,48]. The model has been extensively tested and validated against geologic records from the most recent ~500 million years of Earth's history and can predict the coupled histories and controls on atmospheric $CO_2$, $O_2$ and ocean composition ($PO_4^{3-}$, $NO_3^-$, $SO_4^{2-}$) on geologic timescales. The COPSE model was recently revised by ref.[51], but the version used here is based mainly on a previous version of the model described in ref.[12,15], because we have constructed the model with the aim of keeping the design as simple as possible. We do not anticipate our overarching conclusions to be altered by the differences. Our aim here is to produce a quantitative model that comprehensively encapsulates a range of key biogeochemical processes in oxic/anoxic biospheres and makes resulting predictions for the future lifespan of Earth's oxygenated atmosphere. Specifically, we made the following additions to the basic structure of the model: (1) metabolic $CH_4$ production/consumption processes in the ocean, (2) parameterized atmospheric photochemistry within the $O_2$-$O_3$-$CH_4$ system[24,25], (3) the radiative impact of $CH_4$ on global energy balance[26], (4) a dependence of marine phytoplankton growth rate on surface temperature[24] and marine inorganic carbon availability, (5) exchange of C, S, and Fe between the exogenic (ocean-atmosphere-crust) system and the mantle, and (6) a full redox ($O_2$) budget in the exogenic system.

Primary production and decomposition. In the COPSE model, the activity levels of terrestrial and marine ecosystems are evaluated separately. The global net primary production (NPP), $J_{NPP}$ (in terms of organic carbon), is given by:

$$J_{\text{NPP}} = J_{\text{NPP}}^{\text{ocn}} + J_{\text{NPP}}^{\text{land}}, \tag{1}$$

where $J_{NPP}^{ocn}$ and $J_{NPP}^{land}$ denote marine and terrestrial biospheric productivity, respectively. $J_{NPP}^{ocn}$ is assessed as follows:

$$J_{\text{NPP}}^{\text{ocn}} = f_{\text{n}} \cdot f_{T}^{\text{ocn}} \cdot J_{\text{NPP}}^{\text{ocn},*}, \tag{2}$$

where * denotes the modern value of ~45 Gt C $yr^{-1}$, or 3750 Tmol C $yr^{-1}$ (ref.[10]), and $f_n$ and $f_T^{ocn}$ denote the effect of nutrient availability and surface temperature on oceanic productivity, respectively. A Liebig's minimum law is adopted for $f_n$:

$$f_n = \frac{Minimum\left[ r_{\text{C/N}}N, r_{\text{C/P}}P, M_{\text{CO}_2+\text{HCO}_3^-} \right]}{r_{\text{C/N}}N^*}, \tag{3}$$

1

where N, P, and $M_{CO2+HCO3-}$ denote reservoir sizes of seawater nitrate, phosphate and inorganic carbon ($CO_{2aq}$ and $HCO_3^-$) available for phytoplankton growth, and $r_{C/P}$ (= 117) and $r_{N/P}$ (= 16) denote the "Redfield ratios" of photosynthetic biomass. In the original COPSE model the growth rate of marine phytoplankton is limited by the availability of nutrients (N or P). However, even in the modern ocean, $CO_2$-limitation can reduce growth rates among some phytoplankton taxa[52,53]. Given a secular decrease in atmospheric $CO_2$ levels (Fig. 2c), marine phytoplankton could be limited by the availability of dissolved inorganic carbon ($CO_{2aq}$ and $HCO_3^-$) in the remote future. The dissolved concentrations of $CO_{2aq}$ and $HCO_3^-$ are calculated in the model by solving the seawater carbonate system (see Supplementary Table 3). For the temperature response of marine phytoplankton growth, we adopt the following general function based on ref.[24];

$$f_{\mathrm{T}}^{\mathrm{ocn}}(T_{\mathrm{s}}) = \begin{cases} 0 & (T_{\mathrm{s}} < 273\ \mathrm{K}) \\ k\left\{1-\left(\dfrac{T_{\mathrm{opt}}-T_{\mathrm{s}}}{\Delta T_{\mathrm{opt}}}\right)^2\right\} & (273\ \mathrm{K} \le T_{\mathrm{s}} \le 301\ \mathrm{K}), \\ k\left[\left\{1-\left(\dfrac{T_{\mathrm{opt}}-T_{\mathrm{s}}}{\Delta T_{\mathrm{opt}}}\right)^2\right\}\cdot \exp\left(-\dfrac{\left|T_{\mathrm{s}}-T_{\mathrm{opt}}\right|^3}{T_{\mathrm{ref}}}\right)\right] & (T_{\mathrm{s}} \ge 301\ \mathrm{K}) \end{cases} \tag{4}$$

where k denotes a normalizing factor (= 1.2748), and $T_{opt}$ and $\Delta T_{opt}$ are assumed to be 301 K and 301 $K^3$, respectively. The maximum temperature for phytoplankton growth is controlled by $T_{ref}$, whose uncertainty is explored by the statistical analysis (see below and Extended Data Figs. 1 and 3).

Almost all organic matter produced in the surface ocean is decomposed before burial. Fractions of organic matter decomposed by aerobic respiration (combined with denitrification/nitrification and microbial sulphate reduction/sulphide oxidation) and methanogenesis are expressed as follows:

$$f_{\mathrm{O_2}} = \frac{O}{O + K_{\mathrm{O_2}}}, \tag{5}$$

$$f_{\mathrm{CH_4}} = 1 - f_{\mathrm{O_2}}, \tag{6}$$

where O denotes the oxygen reservoir size in the combined ocean-atmosphere system and $K_{O2}$ = 13.6×10[18] mol. The total organic carbon decomposition rate is given by $J_{\mathrm{NPP}}^{\mathrm{ocn}} - J_{\mathrm{org}}^{\mathrm{b,ocn}} \equiv J_{\mathrm{org}}^{\mathrm{r,ocn}}$, where $J_{org}^{b,ocn}$ denotes the burial rate of organic carbon in marine sediments that is proportional to the square of the normalized primary production as in the original COPSE model[48]. The global methane generation rate via methanogenesis ($2CH_2O \rightarrow CH_4 + CO_2$) is given by $0.5 \times f_{\mathrm{CH_4}} \times J_{\mathrm{org}}^{\mathrm{r,ocn}}$. A fraction of this methane is consumed by aerobic/anaerobic methane oxidation processes (aerobic methanotrophy and AOM), given by:

$$\delta = \frac{O}{O + K_{\mathrm{O_2}}'}, \tag{7}$$

where $K_{O2}'$ = 0.273×10[18] mol (ref. [54]). Given the stoichiometric relationships, the net production rate of $CH_4$, $CO_2$, and $O_2$ in the ocean interior are given by:

$$J_{\mathrm{CH_4}}^{\mathrm{ocn}} = \frac{1}{2}(1-\delta)\cdot f_{\mathrm{CH_4}} \cdot J_{\mathrm{org}}^{\mathrm{r,ocn}}, \quad (8)$$

$$J_{\mathrm{CO_2}}^{\mathrm{ocn}} = \left( f_{\mathrm{O_2}} + \frac{1}{2}(1+\delta) f_{\mathrm{CH_4}} \right)\cdot J_{\mathrm{org}}^{\mathrm{r,ocn}}, \quad (9)$$

$$J_{\mathrm{O_2}}^{\mathrm{ocn}} = J_{\mathrm{NPP}}^{\mathrm{ocn}} - \left( f_{\mathrm{O_2}} + \delta f_{\mathrm{CH_4}} \right)\cdot J_{\mathrm{org}}^{\mathrm{r,ocn}}. \quad (10)$$

For simplicity, the uncertainty associated with partitioning of $CH_4$ consumption between aerobic methanotrophy and AOM is accounted for implicitly by prescribing $K_{O2}$'.

The NPP of Earth's terrestrial biosphere is scaled by a factor V which represents global land biomass normalized by the modern value[48]:

$$J_{\mathrm{NPP}}^{\mathrm{land}} = V \cdot J_{\mathrm{NPP}}^{\mathrm{land,*}}, \quad (11)$$

where the NPP of the present terrestrial biosphere, $J_{NPP}^{land,*}$, is ~60 Gt Cyr$^{-1}$, or 5000 Tmol C yr$^{-1}$ (ref.[10]), and V is calculated, as follows:

$$V = f_{\mathrm{UV}}(\mathrm{O_2})\cdot f_{\mathrm{P}} \cdot f_{\mathrm{fire}}(\mathrm{O_2})\cdot V_{\mathrm{npp}}(\mathrm{O_2},\mathrm{CO_2},T), \quad (12)$$

where $V_{npp}$ denotes the effects of atmospheric $O_2$ and $CO_2$ concentrations and surface temperature on terrestrial biomass as in the original model (Extended Data Fig. 1a,b)[48], $f_{fire}$ denotes the effect of the fire feedback[12,48], and $f_P$ denotes a forcing factor representing selective-P weathering by vascular plant activity[12,15]. Our model introduces an additional factor, $f_{UV}$, in order to capture the effect of UV on the terrestrial biosphere as a function of atmospheric $O_2$ levels:

$$f_{\mathrm{UV}}(\mathrm{O_2}) = \tanh\left( \frac{p\mathrm{O_2}(\mathrm{PAL})}{c_{\mathrm{UV}}} \right), \quad (13)$$

where $pO_2$(PAL) denotes atmospheric $O_2$ levels in terms of PAL, and $c_{UV}$ is treated as one of the model parameters explored during our stochastic analysis (see below). The minimum $CO_2$ concentration for the terrestrial plant activity is also treated as one of the uncertainties, because the ecological $CO_2$ threshold is not precisely known and its absolute value is likely to change with global mean temperature via its influence on photosynthesizer respiration rate[55,56].

The $CH_4$ flux from the terrestrial biosphere is assumed to be proportional to the burial rate of organic matter on land, as follows:

$$J_{\mathrm{CH_4}}^{\mathrm{land}} = \left( J_{\mathrm{org}}^{\mathrm{b,land}} / J_{\mathrm{org}}^{\mathrm{b,land*}} \right)\cdot f_T^{\mathrm{land}} \cdot J_{\mathrm{CH_4}}^{\mathrm{land,*}}, \quad (14)$$

where $f_T^{land}$ denotes the temperature effect on the $CH_4$ flux[57]:

$$f_T^{\text{land}} = a_{\text{CH}_4} \cdot \exp\left(b_{\text{CH}_4} \cdot T_{\text{C}} / T_{\text{C}}^*\right), \quad (15)$$

where $T_C$ denotes the surface temperature in terms of $^{o}$C. On the other hand, the net production rate of $CH_4$, $CO_2$, and $O_2$ in the terrestrial biosphere can be written, as follows:

$$J_{\text{CH}_4}^{\text{land}} = \frac{1}{2}(1-\delta) \cdot g_{\text{CH}_4} \cdot J^{\text{r,land}}, \quad (16)$$

$$J_{\text{CO}_2}^{\text{land}} = \left(g_{\text{O}_2} + \frac{1}{2}(1+\delta) g_{\text{CH}_4}\right) \cdot J^{\text{r,land}}, \quad (17)$$

$$J_{\text{O}_2}^{\text{land}} = J_{\text{NPP}}^{\text{land}} - \left(g_{\text{O}_2} + \delta g_{\text{CH}_4}\right) \cdot J^{\text{r,land}}, \quad (18)$$

where $J^{r,land}$ ($= J_{NPP}^{land} - J_{org}^{b,land}$) denotes the decomposition rate of organic matter in the terrestrial biosphere, and the fraction of methane oxidation is calculated by equation (7) above. A fraction of organic matter decomposed by methanogenesis, $g_{CH4}$, is determined from equations (14) and (16). Then $g_{O2}$ is determined from $g_{\text{O}_2} = 1 - g_{\text{CH}_4}$.

**Photochemistry.** The revised model includes parameterized photochemistry that allows calculation of the concentrations of atmospheric $O_2$ and $CH_4$. We incorporate a scheme for parameterized $O_2$-$O_3$-$CH_4$ photochemistry based on the previous study of a 1-D photochemical model[24]. Specifically, the $CH_4$ oxidation rate $k_{CH4ox}$ (in mol$^{-1}$ y$^{-1}$) is expressed as a polynomial function of the reservoir sizes of $O_2$ and $CH_4$ (O and M)[25]:

$$\log k_{\text{CH}_4\text{ox}} = \alpha_0^j + \alpha_1^j \cdot \varphi_{\text{O}_2} + \alpha_2^j \cdot \varphi_{\text{O}_2}{}^2 + \alpha_3^j \cdot \varphi_{\text{O}_2}{}^3 + \alpha_4^j \cdot \varphi_{\text{O}_2}{}^4 + \alpha_5^j \cdot \varphi_{\text{O}_2}{}^5 + \alpha_6^j \cdot \varphi_{\text{O}_2}{}^6, \quad (19)$$

where $\alpha_i^j$ is fitting coefficients for a given atmospheric $CH_4$ levels and $\varphi_{O2}$ is given by $\log pO_2$(bar) (Supplementary Table 4). The oxidation rate is evaluated based on Fig. 3 of ref. [24] showing the oxidation rate as a function of $pO_2$ and $pCH_4$. We took the relationship between $k_{CH4ox}$ and $pO_2$ for $pCH_4$ of $10^{-6}$, $10^{-5}$, $10^{-4}$, $10^{-3}$, $2\times10^{-3}$ bar, and $k_{CH4ox}$ is calculated as a function of $pO_2$ and $pCH_4$ with a log-linear interpolation method.

For the atmospheres examined in this study, $CH_4$ and $H_2O$ are the dominant hydrogen-bearing species in the stratosphere, allowing the hydrogen escape flux, $J_{esc}$, to be calculated as:

$$J_{\text{Hesc}} = f_{\text{esc}} \cdot f_{\text{T}}(\text{H}_2) = J_{\text{Hesc}}^{\text{H}_2\text{O}} + J_{\text{Hesc}}^{\text{CH}_4}, \quad (20)$$

where the proportional coefficient, $f_{esc}$, is $6.68\times10^{15}$ mol yr$^{-1}$ and $f_T(H_2)$ denotes the total hydrogen molecule mixing ratio:

$$f_{\text{T}}(\text{H}_2) = f\text{H}_2\text{O} + 2 f\text{CH}_4. \quad (21)$$

Therefore, fluxes of hydrogen escape via $CH_4$ and $H_2O$ are written in terms of the abundances in the stratosphere as:

$$\begin{aligned} J_{\mathrm{Hesc}}^{\mathrm{H_2O}} &= f_{\mathrm{esc}} \cdot f\mathrm{H_2O} \\ J_{\mathrm{Hesc}}^{\mathrm{CH_4}} &= 2 \cdot f_{\mathrm{esc}} \cdot f\mathrm{CH_4}. \end{aligned} \tag{22}$$

We fit the empirical relationship between the stratospheric water vapour concentration and global average surface temperature obtained from a suite of 3-D GCM simulations of ref. [35] with a 3rd order polynomial in surface temperature:

$$\log_{10} f\mathrm{H_2O} = f_0 + f_1 T_K + f_2 {T_K}^2 + f_3 {T_K}^3, \tag{23}$$

where $f_i$ denotes the fitting coefficients ($f_0$ = 198.34, $f_1$ = -1.6856, $f_2$ = 4.147×10$^{-3}$, and $f_3$ = -2.7947×10$^{-6}$) (Supplementary Fig. 2).

**Subduction flux.** Despite its crucial role in controlling the redox budget in the exogenic system, a process not typically represented in large-scale biogeochemical models is the subduction of redox sensitive elements (C, S, and Fe species) into the mantle. The exchange flux between Earth surface and the mantle influences not only the sizes of the surficial reservoirs but the oxygen balance of the atmosphere on the long timescales considered here. For the modern Earth the subduction flux of organic C is estimated at <0.2-0.7 Tmol $O_2$ equiv. yr$^{-1}$ (ref. [40,42]), and that of sedimentary pyrite would also be small (<1 Tmol S yr$^{-1}$)[42,58]. On the other hand, the rate of subduction of ferric iron could be 3 Tmol Fe yr$^{-1}$ (0.75 Tmol $O_2$ equiv. yr$^{-1}$)[40], or as high as 7 Tmol Fe yr$^{-1}$ (1.75 Tmol $O_2$ equiv. yr$^{-1}$)[59] to 19 Tmol Fe yr$^{-1}$ (4.8 Tmol $O_2$ equiv. yr$^{-1}$)[42]. However, there is a large uncertainty in the major oxidants ($O_2$ vs. $SO_4^{2-}$) for $Fe^{2+}$ during aqueous oxidation of seafloor[60]. These redox-dependent subduction fluxes probably vary in intensity in response to changes in tectonic activity and oceanic redox state. In this study, the subduction fluxes of C, S, and Fe are parameterized as follows:

$$J_{\mathrm{org}}^{\mathrm{s}} = J_{\mathrm{org}}^{\mathrm{s,*}} \cdot \left( \frac{J_{\mathrm{org}}^{\mathrm{b}}}{J_{\mathrm{org}}^{\mathrm{b,*}}} \right) \cdot f_{\mathrm{G}}, \tag{24}$$

$$J_{\mathrm{carb}}^{\mathrm{s}} = J_{\mathrm{carb}}^{\mathrm{s,*}} \cdot \left( \frac{J_{\mathrm{carb}}^{\mathrm{b}}}{J_{\mathrm{carb}}^{\mathrm{b,*}}} \right) \cdot f_{\mathrm{G}} \tag{25}$$

$$J_{\mathrm{py}}^{\mathrm{s}} = J_{\mathrm{py}}^{\mathrm{s,*}} \cdot \left( \frac{J_{\mathrm{py}}^{\mathrm{b}}}{J_{\mathrm{py}}^{\mathrm{b,*}}} \right) \cdot f_{\mathrm{G}} \tag{26}$$

$$J_{\mathrm{Fe\text{-}ox}}^{\mathrm{s}} = J_{\mathrm{Fe\text{-}ox}}^{\mathrm{s,*}} \cdot \left( \frac{1 - DOA}{1 - DOA^*} \right) \cdot f_{\mathrm{G}} \tag{27}$$

where * denotes a reference value, which is treated as a free parameter. The rate of subduction is scaled to the tectonic activity represented by a forcing $f_G$ (D in the original COPSE model). The subduction fluxes of organic carbon and carbonate carbon are also proportional to their burial fluxes. We also assume that the subduction flux of Fe-oxides is a function of the degree of anoxia (DOA) which represents oceanic anoxicity. The rate of subduction of ferric iron is thought to be sensitive to the redox landscape of the global ocean (e.g., sulfidic or Fe-rich anoxia), but this is outside the scope of the present work. Even if subduction of iron oxides is assumed to continue in an anoxic, Fe-rich ocean, this would not change our overarching conclusion that the future lifespan of the oxygenated atmosphere is less than 1.5 Gyr because the subduction of ferric iron represents the loss of oxidizing power from the surface system. Given the large uncertainty in the magnitude of subduction fluxes, we explore a wide range of reference values in our stochastic analysis (Extended Data Fig. 3).

**Global $O_2$ budget.** In this study, redox balance in the model is tracked in units scaled to the oxidizing power of $O_2$. The model explicitly accounts for fluxes of oxidants and reductants into and out of the surface system (Supplementary Fig. 3). The global redox budget (GRB) for the ocean-atmosphere system (OAS) is written as follows:

$$GRB_{\mathrm{OAS}} = 0.5J_{\mathrm{Hesc}} + \left(J_{\mathrm{org}}^{\mathrm{b,ocn}} + J_{\mathrm{org}}^{\mathrm{b,t}} - J_{\mathrm{org}}^{\mathrm{w}} - J_{\mathrm{org}}^{\mathrm{m}}\right) + \left(2J_{\mathrm{py}}^{\mathrm{b}} - 2J_{\mathrm{py}}^{\mathrm{w}}\right) - 2J_{\mathrm{H_2S}}^{\mathrm{d}} - J_{\mathrm{red}}^{\mathrm{d}} - 0.25J_{\mathrm{Fe\text{-}ox}}^{\mathrm{s}} \quad (28)$$

where $J_{\mathrm{Hesc}}$ denotes the irreversible hydrogen escape to space, and second and third terms represent the influence of organic C and pyrite S subcycles on the GRB. $J_{\mathrm{H2S}}{}^{\mathrm{d}}$ and $J_{\mathrm{red}}{}^{\mathrm{d}}$ are the mantle degassing of $H_2S$ and other reducing gases (e.g., $H_2$). $J_{\mathrm{Fe\text{-}Ox}}{}^{\mathrm{s}}$ is the subduction flux of ferrous iron. In this study, we also revised the baseline COPSE model to capture the $O_2$ budget in the exogenic (ocean-atmosphere-crust) system on geological timescales, according to:

$$GRB_{\mathrm{exo}} = 0.5J_{\mathrm{Hesc}} + J_{\mathrm{org}}^{\mathrm{s}} + 2J_{\mathrm{py}}^{\mathrm{s}} - 0.25J_{\mathrm{Fe\text{-}ox}}^{\mathrm{s}} - 2J_{\mathrm{H_2S}}^{\mathrm{d}} - J_{\mathrm{red}}^{\mathrm{d}}. \quad (29)$$

The primary sink of reducing power from the exogenic system is the irreversible escape of hydrogen to space ($J_{\mathrm{Hesc}}$) and the subduction of organic C and pyrite S to the mantle ($J_{\mathrm{org}}{}^{\mathrm{s}}$ and $J_{\mathrm{py}}{}^{\mathrm{s}}$), while the primary source of reducing power is outgassing of reduced gases from the Earth's interior ($J_{\mathrm{H2S}}{}^{\mathrm{d}}$ and $J_{\mathrm{red}}{}^{\mathrm{d}}$). The subduction of ferric iron ($J_{\mathrm{Fe\text{-}ox}}{}^{\mathrm{s}}$) is also a loss of oxidizing power (i.e., source of reducing power). The net input flux of reducing power from the mantle to the exogenic system is defined as:

$$\Phi(\mathrm{Red}) = -J_{\mathrm{org}}^{\mathrm{s}} - 2J_{\mathrm{py}}^{\mathrm{s}} + 0.25J_{\mathrm{Fe\text{-}ox}}^{\mathrm{s}} + 2J_{\mathrm{H_2S}}^{\mathrm{d}} + J_{\mathrm{red}}^{\mathrm{d}}. \quad (30)$$

In Fig. 3b of the main text, we assess the impact of long-term Φ(Red) on the future lifespan of oxygenated atmosphere. To do this, we average Φ (Red) over the future lifespan:

$$\overline{\Phi_i(\mathrm{Red})} = \frac{\int_0^{\tau_i} \Phi(\mathrm{Red})dt}{\tau_i}, \quad (31)$$

where $\tau_i$ denotes the future lifespan for the different values of $O_2$ threshold i (in PAL). Existing empirical estimates of the mantle degassing of reducing gases via submarine volcanism, 0.7±0.3 Tmol $O_2$ equiv. $yr^{-1}$ (ref. [60]), the limited $O_2$ consumption through seafloor $Fe^{2+}$ oxidation 0.14 Tmol $O_2$ equiv. $yr^{-1}$ (ref. [60]), and the subduction flux of organic C of <0.2-0.7 Tmol $O_2$ equiv. yr-

[1], yield a present net input flux of reducing power from the mantle to the surface system, □(Red)*, of >-0.16 Tmol $O_2$ equiv. $yr^{-1}$. A resampling analysis of Φ(Red)* from our successful model simulations predicts a slightly lower value of -0.24±0.27 Tmol $O_2$ equiv. $yr^{-1}$ (Supplementary Fig. 4).

**Climate model.** Global mean surface temperature $T_s$ (in K) is estimated with a zero-dimensional energy balance model (0-D EBM) in which the absorption of solar radiation is balanced by the heat loss by outgoing longwave (i.e., infrared) radiation (OLR) to space ($J_{OLR}$):

$$(1\text{-}\alpha)\cdot\frac{S_{\odot}}{4} = J_{\mathrm{OLR}} - RF_{\mathrm{CH_4}}, \quad (32)$$

where α and $S_{\odot}$ denote the top-of-atmosphere (TOA) albedo and the value of the solar constant at the time under consideration, respectively. For $J_{OLR}$ and α, we adopted the parametric expressions proposed by ref.[61], with the following variables: p = $pCO_2$ (in bar); $\phi = \ln(p/3.3\times10^{-4})$; μ = cos(z), where z is the zenith angle; and surface albedo $\alpha_s$. $\alpha_s$ is written as a function of $T_s$:

$$\alpha_{\mathrm{s}} = \begin{cases} \alpha_{00} & (T_0 \le T_{\mathrm{s}}) \\ \alpha_{\mathrm{ice}} - (\alpha_{\mathrm{ice}} - \alpha_{00})\cdot\dfrac{(T_{\mathrm{s}} - T_{\mathrm{ice}})^2}{(T_{\mathrm{s}} - T_{\mathrm{ice}})^2} & (T_{\mathrm{ice}} \le T_{\mathrm{s}} \le T_0) \\ \alpha_{\mathrm{ice}} & (T_{\mathrm{s}} \le T_{\mathrm{ice}}) \end{cases}, \quad (33)$$

where it is assumed that Earth is globally ice-covered for $T_s < T_{ice}$ (= 263 K) with an albedo $□_{ice}$ (= 0.65), and that Earth is characterized by a relatively low albedo $\alpha_0$ (= 0.35) for $T_s > T_0$ (= 273K). For $T_{ice} < T_s < T_0$, the albedo is interpolated with a quadratic interpolation method[62]. $RF_{CH4}$ represents the radiative forcing by methane, which is quantified based on ref.[26]. Recent 1-D modelling results suggest that relatively small increases in solar luminosity could conceivably trigger a runaway greenhouse on Earth[63,64]. However, 3-D climate models indicate that Earth's climate is likely to be somewhat more stable than this as a consequence of expansion of unsaturated regions in the upper atmosphere attendant to changes in Hadley circulation that occur at higher solar fluxes[23,65]. Given the quantitative consistency between the above 0-D EBM and recent 3-D climate models[35] (Supplementary Fig. 1), we consider our utilization of 0-D EBM a defensible approximation of planetary climate for implementation on $10^6$-$10^9$-year timescales. Nevertheless, it is important to point out that future projections of Earth system behavior will likely be modulated by uncertain cloud feedbacks and continental configurations. As a result, in our MC analysis we explore uncertainties in OLR and TOA (Extended Data Fig. 3).

**Stochastic analysis.** The computational efficiency of the model enables us to examine the future lifespan of the oxygenated atmosphere via a stochastic analysis, in which we randomly sample from a prior distribution for poorly constrained parameters and quantify whole-model uncertainty (Extended Data Fig. 3). We draw poorly constrained initial values for crustal C and S reservoir sizes at 600 Ma (million years ago) over a wide range of values. We also vary the reference values of fluxes between the exogenic system and the mantle (subduction and mantle degassing) to capture uncertainties in model parameterisation. After running the model forwards in time until the present, we only sample model runs that allow for acceptable atmospheric composition ($pO_2$ = 21±1% and $pCO_2$ = 300±150 ppmv), global surface temperature (288±2K) and seawater sulphate ($SO_4^{2-}$) levels ($SO_4^{2-}$ = 25±15 mM). The model runs that fail to obtain reasonable values have been abandoned. For our default stochastic analysis, we performed nearly 400,000 simulations to span the whole parameter space, and the resultant successful runs were n = 4,787 which were continued to be run for estimating the future lifespan of the oxygenated atmosphere.

For the future evolution of the degassing factor ($f_G$) and erosion factor ($f_R$; U in the original model), we assumed following functions:

$$f_G = \left(1 + a_G + a_G \sin\left(2\pi \frac{t - \tau_G/4}{\tau_G}\right)\right) \cdot \left(\left(1 + \frac{t}{4500}\right)^{-n}\right)^m, \quad (34)$$

$$f_R = \left(1 + a_R \sin\left(2\pi \frac{t - \tau_R}{\tau_R}\right)\right) \cdot \left(\left(1 + \frac{t}{4500}\right)^{-n}\right)^m, \quad (35)$$

where the first term on the right-hand side denotes the periodic changes with an amplitude a and a period of $\tau$ (in Myr), and the second term represents the secular decrease in the activity of tectonics[66]. These relationships have been adopted for examining to the geological past, and we consider that this assumption is a reasonable first-order approximation as there is no a priori reason for thinking that this style of decay will not continue into Earth's far future. Nevertheless, we also emphasize that this is very under-studied and it is possible that non-linearity in tectonic regime will be important to consider for Earth and other habitable rocky planets in future work.

We draw model parameters from wide distributions that represent the uncertainty in their values (Extended Data Fig. 3). For example, the uncertainty in the biological responses is assessed by changing biological parameters, such as FERT. The strength of terrestrial weathering feedback is also a topic of debate[67-69], and we explore its uncertainty by changing the apparent activation energy (ACT) and a runoff factor (RUNsil)[70]. The amplification of terrestrial weatherability by land plants is also an important uncertainty[71,72], which we explore here by changing LIFE[70]. To account for the uncertainty in the value of $c_{UV}$, we draw it from a log-uniform distribution between $10^{-2.5}$ PAL and $10^{-1.5}$ PAL (Extended Data Fig. 1c). At the higher end of this range, terrestrial plant activity is suppressed when atmospheric $O_2$ is <10% PAL. At the lower end, the threshold is lowered to ~1%PAL. We consider these two end-member cases to cover the full range of uncertainty of UV impacts on the terrestrial ecosystems. We also explore uncertainty in the strength of redox-dependent P recycling in marine sediments by changing $C_{org}/P_{org}$ ratios for anoxic sediments.

The sampling procedure is run for long enough to obtain stationary distributions for posterior probability densities. However, it is possible that our model predictions may underestimate the inherent variability in the future Earth's evolution. For instance, the possibility of random catastrophe, such as the asteroid impacts, could also introduce additional uncertainty in the estimate of future lifespan of Earth's oxygenated atmosphere. We also note that the uncertainty in the longevity of human impacts on Earth's future environment is beyond the scope of this study.

**Data availability.** The numerical data for Figure 3 in the main text and Extended Data Figures 4 and 5 are presented in the Source data. The big data obtained by the statistical analysis is available to download at https://doi.org/10.6084/m9.figshare.13487487.v1.

**Code availability.** Our Fortran source code is available at https://github.com/kazumi-ozaki/lifespan.